\documentstyle[sprocl,epsfig]{article}

\input{psfig}

\def\be{\begin{equation}}
\def\ee{\end{equation}}
\def\bq{\begin{eqnarray}}
\def\eq{\end{eqnarray}}

\begin{document}

\begin{flushright}
WUE-ITP-99-021
\end{flushright}
\bigskip

\title{PION FORM FACTORS FROM QCD LIGHT-CONE SUM RULES }

\author{A. KHODJAMIRIAN \footnote{ {\it on leave from 
Yerevan Physics Institute, 375036 Yerevan, Armenia }}$^,$\footnote{
after Sept. 1, 1999 at The Niels Bohr Institute,  
DK-2100, Copenhagen, Denmark}}
\address{
Institut f\"ur Theoretische Physik, 
Universit\"at W\"urzburg,\\
D-97074 W\"urzburg, Germany 
\\E-mail : ak@physik.uni-wuerzburg.de}

\maketitle\abstracts {Light-cone sum rules have proved to be 
very useful in calculating hadronic matrix elements for 
exclusive processes. I present recent applications 
of this method to the pion electromagnetic form factor
and to the form factors of  $\gamma^* \rho \to \pi$ 
and $\gamma^*\gamma \to \pi^0$ transitions.\\
\begin{center}
{\it Presented at the 6th INT/Jlab Workshop on Exclusive and 
Semiexclusive Processes at High Momentum Transfer, Jefferson Laboratory, 
May 1999.}
\end{center}}

\section{Introduction}

Current and future experimental studies of exclusive and semi-exclusive 
processes require accurate calculations of relevant 
hadronic matrix elements in QCD. This task 
is, however, still far from being fulfilled. Of particular 
interest are quantitative QCD predictions for 
various form factors  determining matrix elements of 
electroweak quark currents between one-hadron states.
At sufficiently large values of the momentum transfer 
the form factors are determined by the perturbative 
QCD factorization. In particular, the pion electromagnetic 
form factor at $Q^2 \to \infty$ is given by the well known 
expression \cite{CZ,ER,BL}
\begin{equation}
 F_\pi(Q^2) = \frac{8\pi\alpha_s f_\pi^2}{9Q^2}
  \left|\int\limits_0^1\!\! du \frac{\varphi_{\pi}(u)}{\bar u}\right|^2\,,
\label{asympt}
\end{equation}  
$(\bar{u}\equiv 1-u)$ obtained by the convolution of twist-2 distribution 
amplitudes $\varphi_\pi(u)$ of the initial and final pion 
with  the  $O(\alpha_s)$ quark hard-scattering kernel. 
The major unsolved problem is to estimate 
the so-called soft, or end-point contributions to this form factor. 
These contributions are expected \cite{soft} to be important 
at intermediate  momentum transfers $Q^2 \sim 1 \div 10 $ GeV$^2$.
Obviously, a consistent solution of this problem requires 
a calculation of both soft and hard contributions in {\em one 
and the same} framework beyond perturbation theory. 

A promising and largely universal method of   
calculating hadronic matrix elements 
is provided by QCD light-cone sum rules (LCSR) \cite{LCSR}. 
This approach combines the light-cone operator-product expansion 
(OPE) with the conventional SVZ sum rule technique
\cite{SVZ}. Proceeding from the basis of QCD perturbation theory,
LCSR incorporate elements of nonperturbative long-distance dynamics  
parametrized in terms of hadronic distribution amplitudes 
with different twist and multiplicity. 

In this talk, I describe recent applications of LCSR 
to various pion form factors, including  the pion electromagnetic
form factor \cite{BH94,BKM} and the form factors of  
$\gamma^* \rho(\omega)\to \pi$ and $\gamma^* \gamma \to \pi^0$
transitions \cite{AK}. 

\section{Pion electromagnetic form factor}

I begin with explaining the general idea of the LCSR approach 
using an important example of the pion electromagnetic form factor.
More detailed derivation can be found in \cite{BKM}. 
The starting object is the vacuum-pion correlation function 
\begin{equation}
T_{\mu\nu}(p,q) = i\int\! d^4 x \,e^{iqx}
\langle 0| T\{j_\mu^5(0) j_\nu^{\rm em}(x)\} 
| \pi^+(p)\rangle \;,
\label{corr}
\end{equation}
where $j_\nu^{\rm em} = e_u\bar u\gamma_\nu u + e_d\bar d \gamma_\nu d~$
is the quark electromagnetic current. One of the pions is put on-shell 
and the second one  is replaced by the generating current 
$j_\mu^5 = \bar d\gamma_\mu\gamma_5 u$. For the on-shell pions, 
$p^2=m_\pi^2$ vanishes in the chiral limit adopted here. 

At fixed  large $Q^2=-q^2 >> \Lambda_{QCD}^2$
the correlation function (\ref{corr}) is a function of  a single 
invariant $(p-q)^2$. Depending on the value of this variable the amplitude 
$T_{\mu\nu}$ corresponds to different physical pictures.   
At large spacelike $(p-q)^2$, $|(p-q)^2| >> \Lambda_{QCD}^2$,  
the interval $x^2 \to 0$ 
and one is able to employ the light-cone operator expansion 
for the product of two currents in (\ref{corr}). 
The amplitude (\ref{corr}) can then be factorized 
in a convolution of the pion distribution amplitude and the
hard scattering amplitude yielding 
\begin{equation}
  T_{\mu\nu} =  2if_\pi p_\mu p_\nu \int\limits_0^1 \! du\! 
   \frac{u\varphi_\pi(u)}{\bar{u}Q^2 -u(p-q)^2}+\ldots\,.
\label{ope}
\end{equation}   
In the above, only the leading order, 
twist 2 contribution is shown, the ellipses denoting
$O(\alpha_s)$ corrections and higher twist terms.  

Decreasing $|(p-q)^2 |$, one gradually 
approaches the physical region in the channel of the pion current.
The interval $x^2$ deviates from zero, the light-cone 
OPE diverges and factorization is lost. 
Finally, at $(p-q)^2 = m_\pi^2$ one deals with the 
emission and propagation of an on-shell pion which is then 
elastically scattered by the electromagnetic current.
In this region, the amplitude $T_{\mu\nu}$ is dominated by 
the one-pion contribution to its hadronic representation 
(dispersion relation) 
\begin{equation}
T_{\mu\nu}(p,q) =  2if_\pi (p-q)_\mu p_\nu F_\pi(Q^2)\frac{1}
{m_\pi^2 - (p-q)^2} + \int ds ~\frac{\rho_{\mu\nu}(s)}{
s-(p-q)^2}\, .  
\label{pion}
\end{equation}
The one-pion term is proportional to the 
pion decay constant $f_\pi$ and to the desired form 
factor $F_\pi(Q^2)$.
The integral over $\rho_{\mu\nu}$ in (\ref{pion})
contains contributions 
of higher mass intermediate states with the pion quantum numbers. 

Equating (\ref{ope}) and (\ref{pion})
one extracts  the form factor $F_\pi(Q^2)$ 
applying the standard elements of the QCD sum rule technique.
Specifically, the quark-hadron duality is used: 
$\rho_{\mu\nu}(s) = 1/\pi\mbox{Im}T_{\mu\nu}(s)\Theta(s-s_0^\pi)$,
where $\mbox{Im}T_{\mu\nu}(s)$ is calculated from 
(\ref{ope}) and $s_0^\pi\simeq 0.7$ GeV$^2$ is the effective threshold 
for the pion channel determined from the two-point QCD sum rule  
\cite{SVZ}. Furthermore, the Borel transformation 
in the variable $(p-q)^2$ is performed.  
The resulting LCSR, in the zeroth order 
in $\alpha_s$ and  in the twist 2 approximation, reads \cite{BH94}
\begin{equation}
F_{\pi}(Q^2) = \int\limits_{u_0}^1 \!du\, \varphi_{\pi}(u,\mu_u) 
\exp \left( - \frac{\bar uQ^2}{u M^2} \right)
\stackrel{Q^2\to\infty}{\longrightarrow}
\varphi_{\pi}'(0) 
\int\limits_0^{s_0} \frac{ds\, s\, e^{-s/{M^2}}}{Q^4}\,, 
\label{SR1}
\end{equation} 
where $\varphi_{\pi}'(0)=-\varphi_{\pi}'(1)$, $M^2$ is the Borel 
parameter, $u_0= Q^2/(s_0^\pi+Q^2)$.
The factorization scale $\mu_u^2= \bar{u}Q^2+uM^2 $  
corresponds to the quark virtuality in the correlation function. 
This sum rule perfectly behaves at $Q^2 \to \infty$, in contrast
to the conventional QCD sum rule \cite{IofSmilRad} for $F_\pi(Q^2)$ 
based on the local OPE. 
The  $1/Q^4$ behaviour of (\ref{SR1}) corresponds to the 
soft end-point mechanism, provided that 
in $Q^2\to \infty$ limit 
the integration region in (\ref{SR1}) shrinks to a point $u=1$.

Perturbative $O(\alpha_s)$ corrections to the correlation
function (\ref{corr}) considerably improve  
the accuracy of the soft contribution. More importantly,
in $O(\alpha_s)$ one recoveres the  $\sim 1/Q^2$ asymptotic 
behaviour  corresponding to the hard perturbative mechanism.
The $O(\alpha_s)$ twist 2 term was calculated in \cite{BKM}.
Including this contribution in the LCSR and retaining the 
first two terms of the sum rule expansion in powers 
of $1/Q^2$ one obtains:
$$
F_{\pi} (Q^2) = 
\frac{\alpha_s}{2\pi}C_F\int\limits_0^{s_0} \frac{ds\,  e^{-s/{M^2}}}{Q^2} 
\int\limits_0^1 du\, \frac{\varphi_{\pi}(u)}{\bar u}
+\varphi_{\pi}'(0) \int\limits_0^{s_0} \frac{ds\, s\, e^{-s/{M^2}}}{Q^4}
$$
$$
+
\frac{\alpha_s}{4\pi}C_F \int\limits_0^{s_0} \frac{ds\, s\, e^{-s/{M^2}}}{Q^4}
\Bigg\{ \varphi_{\pi}'(0) \left[-9 + \frac{1}{3}\pi^2 + 
\ln\frac{s}{\mu^2} - \ln^2\frac{s}{Q^2}\right]
$$
\begin{equation}
+(2\ln\frac{s}{\mu^2}-3)\int\limits_0^1 du 
 \left[ \frac{\varphi_{\pi}(u)-\bar u\,\varphi'(0)}{\bar u^2}\right]
+ (2 \ln\frac{s}{\mu^2}-8) \int\limits_0^1\!\! du  
\frac{\varphi_{\pi}(u)}{\bar u}\Bigg\}.
\label{pert}
\end{equation}
It is remarkable that the leading asymptotic $O(1/Q^2)$ term in (\ref{pert})
coinsides with (\ref{asympt}) provided that  
the two-point sum rule \cite{SVZ} for $f_\pi$ 
yields  $\int_0^{s_0}\!ds\,e^{-s/M^2}= 4\pi^2f_\pi^2$,
and $\int_0^1 \!\!du\varphi^{\rm as}_{\pi}(u)/\bar{u} =3$ 
for the asymptotic  $\varphi^{\rm as}_{\pi}(u)=6u\bar{u}$. 
Furthermore, it is instructive to split the $0<u<1$ integration region 
in LCSR into ``hard''  $( u<u_0)$ and ``soft'' $( u>u_0)$ parts.
This separation reveals \cite{BKM} that the $O(\alpha_s/Q^4)$ 
hard contribution is large and negative while the soft  $O(1/Q^4)$ and 
$O(\alpha_s/Q^4)$ contributions  are positive yielding 
considerable cancellations in the sum rule.

To further improve the accuracy of the LCSR, one also has to include
the higher twist corrections. Physically, 
these corrections take into account 
both the transverse momentum of the quark-antiquark state and 
the contributions of higher Fock states in the pion wave function.
In addition to the twist 4 terms, in \cite{BKM} the factorizable 
twist 6 contributions proportional to the square 
of the quark condensate have been calculated. The latter
corrections turn out to be comfortably small.
\begin{figure}[t]
\vspace{-1.2cm}
\hspace{2cm}
\psfig{figure=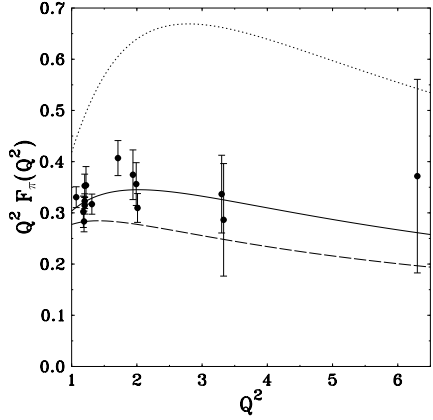,height=2.7in}
\vspace{-0.5cm}
\caption{ The light-cone sum rule predictions for the pion electromagnetic 
  form factor $^8$ using asymptotic distribution amplitude 
   (dashed), CZ distribution (dotted) and fit to the data $^{11}$ (solid).}
\vspace{-0.5cm}
\end{figure}
Adding twist 4,6 terms to the twist 2 (leading and $O(\alpha_s)$) parts 
yields the LCSR prediction for $F_\pi(Q^2)$  shown in Fig. 1 for 
the pion distribution amplitude $
\varphi_\pi(u,\mu) = 6u\bar{u} \left[1+ a_2(\mu)C_2^{3/2}(u-\bar{u})\right]\, 
$
with two choices: $a_2=0$ (asymptotic) and $a_2( 1\,\mbox{GeV})= 2/3 $ (CZ). 
The  available experimental data \cite{exp} 
seem to rule out the CZ option. The fit of the LCSR 
to these data yields $a_2( 1 \,\mbox{GeV} ) = 0.12 \pm 0.07 
^{+0.05}_{-0.07}$ where the first (second) uncertainty is 
experimental (theoretical).
\begin{figure}[t]
\vspace{-1.2cm}
\hspace{2cm}
\psfig{figure=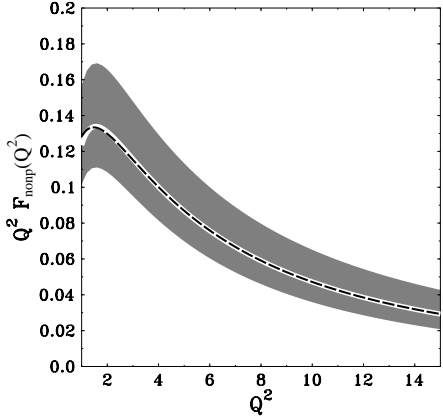,height=2.7in}
\vspace{-0.5cm}
\caption{ The light-cone sum rule prediction$^8$ for the nonperturbative correction
to the pion form factor  with the asymptotic pion distribution amplitude.
The grey band shows the sensitivity of the result to variation of the  
Borel parameter within $0.8 ~{\rm GeV}^2 < M^2 < 1.5 ~{\rm GeV}^2$.
The dashed curve is the calculation for the nominal value 
$M^2 = 1 {\rm GeV}^2$.}
\vspace{-0.3cm}
\end{figure}

Another way to employ the LCSR prediction \cite{BKM}
is to  subtract the leading asymptotic $1/Q^2$ term (the first term 
in (\ref{pert})) from the sum rule. The remaining power suppressed  
part can be  called ``nonperturbative'' contribution $F_{nonp}(Q^2)$. 
One then obtains  a quantitative prediction 
for the full pion form factor by adding $F_{nonp}(Q^2)$ to the 
perturbative QCD result which is currently known 
with the NLO accuracy. The LCSR prediction for $F_{nonp}(Q^2) $ 
is plotted in Fig. 2.  It turns out to be numerically 
moderate due to abovementioned cancellations  between soft 
and hard $O(1/Q^4)$  contributions in the twist 2 part.

\section{ $\gamma^* \rho \to \pi$ and $\gamma^* \gamma \to \pi$
transition form factors} 

Similar to the pion electromagnetic form factor, the 
$\gamma^* \rho \to \pi $ or $\gamma^* \omega \to \pi $ 
transition form factors can be measured by extracting the one-pion 
exchange in the electroproduction of $\rho$ or $\omega$.
The form factors are defined as 
$ \frac13\langle \pi^0(p)\mid j_\mu^{em} \mid \omega(p-q)\rangle \simeq
\langle \pi^0(p)\mid j_\mu^{em} \mid \rho^0(p-q)\rangle =
F^{\rho \pi}(Q^2)m_\rho^{-1}\epsilon_{\mu\nu\alpha\beta}
e^{\nu}q^\alpha p^\beta 
\,,
$
$e_\nu$ being the polarization vector of $\rho$ or $\omega$.
The derivation of LCSR \cite{AK} for $F^{\rho \pi}(Q^2)$ 
basically follows the procedure described in the previous section. 
The underlying correlation function is  
\be
\int \! d^4x e^{-iqx}\langle \pi^0 (p)\! \mid
T\{ j_\mu^{em}(x) j_\nu^{em}(0)\}\mid \! 0\rangle = 
i\epsilon_{\mu\nu\alpha\beta}
q^\alpha p^\beta F^{\gamma^{*}\pi}(Q^2,(p-q)^2).
\label{ampl}
\ee
If both $Q^2$ and $|(p-q)^2|$ are sufficiently large,  
the light-cone OPE is valid for the amplitude 
$F^{\gamma^{*}\pi}(Q^2,(p-q)^2)$ starting from the leading twist 
2 term \cite{BL}: 
\be
F^{\gamma^{*}\pi}(Q^2,(p-q)^2)= \frac{\sqrt{2}f_\pi}{3}\int\limits_0^1 
\frac{du~\varphi_\pi(u)}{\bar{u}Q^2 -u(p-q)^2}~ + ...
\label{leading}
\ee
Physical states in the $(p-q)^2$--channel include  
vector mesons $\rho,\omega,\rho',\omega',...$ and a continuum
of hadronic states with the same quantum numbers. 
The hadronic dispersion relation can be written as 
\be
F^{\gamma^{*}\pi}(Q^2,(p-q)^2) 
=\frac{\sqrt{2}f_\rho F^{\rho \pi}(Q^2)}{m_{\rho}^2-(p-q)^2}
+ \frac{1}{\pi}\int\limits_{s_0^\rho}^\infty ds~ 
\frac{\mbox{Im}F^{\gamma^{*}\pi}(Q^2,s)}{s-(p-q)^2}.  
\label{disp}
\ee
Here, the $\rho$ and $\omega$  contributions
are combined in one ground-state resonance term 
using $m_\rho \simeq m_\omega$ and 
$
3\langle \omega \mid j_\nu^{em} \mid 0 \rangle \simeq
\langle \rho^0 \mid j_\nu^{em} \mid 0 \rangle = 
(f_\rho/\sqrt{2}) m_\rho e^{ *}_\nu  
$. For the higher states above the effective threshold $s_0^\rho$, the
quark-hadron duality is used 
with the spectral density $1/\pi\mbox{Im}F^{\gamma^{*}\pi}(Q^2,s)$ 
calculated from (\ref{leading}). The resulting LCSR 
\be
F^{\rho\pi}(Q^2) = \frac{f_\pi}{3f_\rho} 
\int^1_{u_0}\frac{du}u \varphi_\pi (u)
\exp\left(-\frac{Q^2(1-u)}{uM^2} +\frac{m_\rho^2}{M^2}\right)
+ O(\mbox{twist 4}) 
\label{rhopi}
\ee
was obtained in \cite{AK} where one can 
find additional details and numerical predictions.  
Note that the form factor (\ref{rhopi})
corresponds to the  soft mechanism with $1/Q^4$ behaviour
at large $Q^2$. Due to helicity suppression, the perturbative 
$O(\alpha_s)$ correction to $F^{\rho\pi}(Q^2)$ should have the same 
$1/Q^4$ behaviour. This and twist 6 corrections 
still have to be included in (\ref{rhopi}).

Finally, let me focus on the $\gamma^*\gamma \to  \pi$ transition. 
Since the real photon is a  large-distance  object, 
the transition form factor  $F^{\gamma\pi}(Q^2)$ 
contains nonperturbative contributions beyound the light-cone
expansion of two electromagnetic quark currents. 
The leading $1/Q^2$ asymptotics of $F^{\gamma\pi}(Q^2)$ 
is well known \cite{BL} and given by (\ref{leading})
at $(p-q)^2\to 0$. The calculation of the contributions 
suppressed by powers of $1/Q^2$  is a nontrivial task. 
Within LCSR approach, one has to invoke the photon 
distribution amplitudes. A reliable estimate is nevertheless
possible \cite{AK} using the hadronic dispersion relation (\ref{disp})
where the resonance term is determined by the LCSR (\ref{rhopi})
and the integral over higher states is estimated using duality.  
One can analytically continue  this relation 
to $(p-q)^2 \to 0$ provided that it does not contain  
subtraction terms (a similar approach was used in \cite{GIKO} for 
the structure function of the real photon).
The result shown in Fig. 3 again has a better agreement 
with the experimental data \cite{CELLO} in the case of 
the asymptotic pion distribution amplitude. The recent update \cite{SY} of  
$F^{\gamma\pi}(Q^2)$ includes the 
$O(\alpha_s)$ correction decreasing the leading order results in Fig. 3 by  
15-20\%. 
\begin{figure}[t]
\vspace{3.4cm}
\psfig{figure=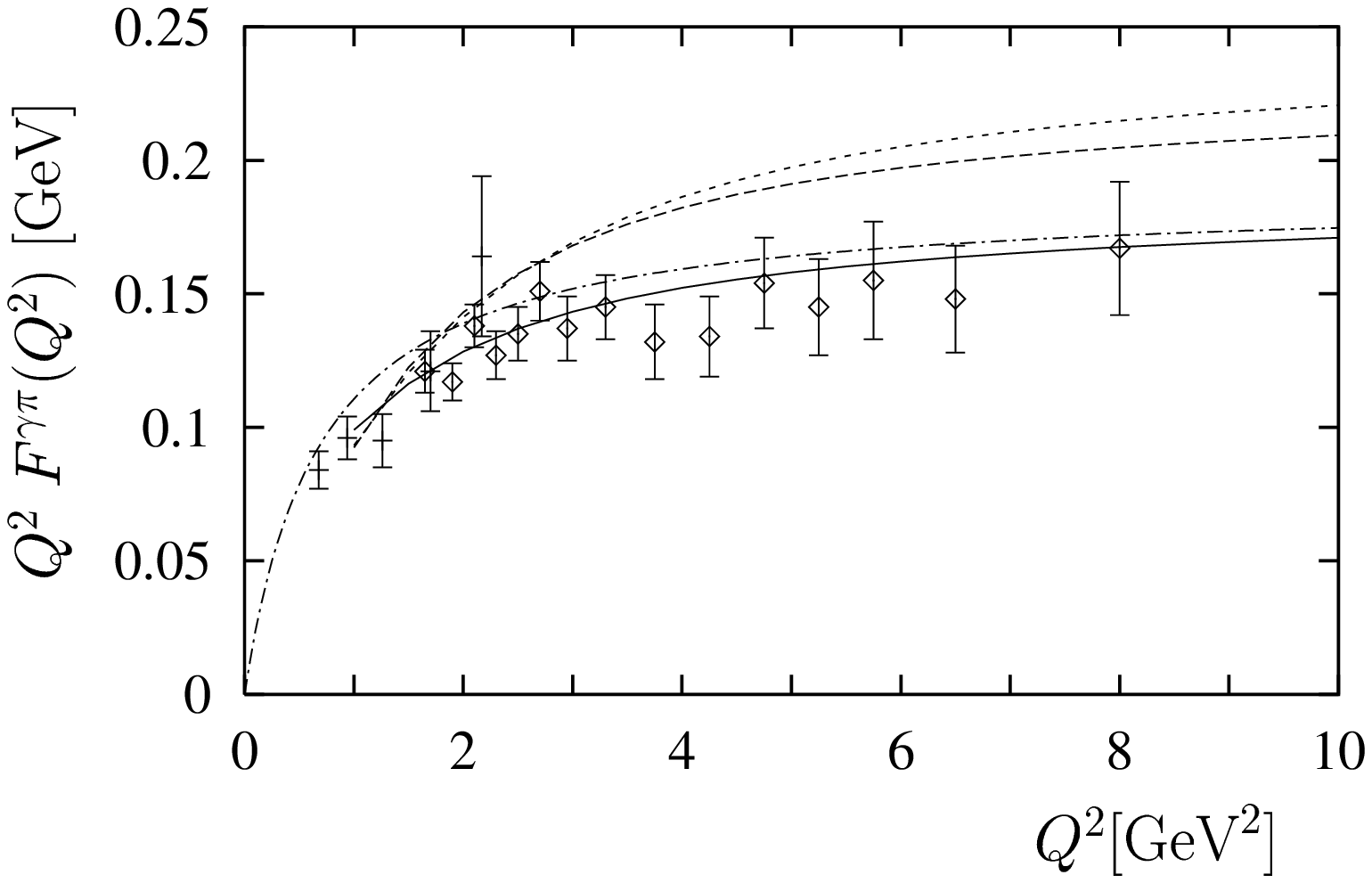,height=2.0in}
\vspace{-4cm}
\caption{Form factor of the $\gamma^*\gamma \to \pi^0$ transition 
calculated $^9$ with the asymptotic 
(solid), CZ (long dashed) and BF (short-dashed) wave 
function of the pion in comparison with the experimental 
data points $^{13}$ and with the  
interpolation formula from $^3$ (dash-dotted).}
\vspace{-0.4cm}
\end{figure}

\section{Conclusions}

Light-cone sum rules provide a powerful tool for calculating various
form factors at intermediate momentum transfers. In this framework, 
both soft end-point and hard perturbative mechanisms 
can be incorporated. The LCSR results for the pion electromagnetic 
form factor  indicate that the soft contribution is indeed large. 
However, an essential part of it is cancelled by the 
$O(\alpha_s/Q^4)$ corrections originating from the hard region. 
As a result, the overall power suppressed  correction to 
the perturbative QCD factorization turns out to be relatively moderate.
Comparison of LCSR predictions for the pion 
form factors with the available data indicates  an almost 
asymptotic pion distribution amplitude $\varphi_\pi(u)$. 
More precise measurements of the form factors 
are needed to tightly constrain the nonasymptotic parts of
$\varphi_\pi(u)$. 

The LCSR approach 
is extremely useful also for heavy flavour exclusive  decays, 
in particular, for the form factors
of $B\to \pi,K,\rho,K^*$ transitions \cite{heavy}. 
The sum rules for heavy-to-light and light-to-light 
transition form factors share a common nonperturbative input,
that is the set of light hadron distribution amplitudes.
Moreover, the structure of 
perturbative corrections to the pion form factor \cite{BKM} 
turns out to be  very similar to the 
heavy quark limit \cite{Bagan} of the LCSR for $B\to \pi$ in 
$O(\alpha_s)$. One may conclude 
that within a universal approach of QCD 
light-cone sum rules, both fields, the hard exclusive 
processes and the heavy flavour decays  can mutually benefit.

\section*{Acknowledgments}

I am grateful to C. Carlson and A. Radyushkin for the invitation
to this fruitful workshop. Collaboration with V. Braun 
and M. Maul on the subject of this talk is acknowledged. 
This work was supported by the German Federal Ministry for 
Education and Research (BMBF) under contract number 05 7WZ91P (0).


\section*{References}

\begin{thebibliography}{99}

\bibitem{CZ}
 V.L.\ Chernyak and A.R.\ Zhitnitsky, JETP Lett.\ 
  {\bf {25}} (1977) 510; Yad.\ Fiz.\ {\bf 31} (1980) 1053;
  Phys. Rept. {\bf 112} (1984) 173.
\bibitem{ER}
  A.V.\ Efremov and A.V.\ Radyushkin, Phys.\ Lett.\ B {\bf 94} (1980)
  245;
  Theor.\ Math.\ Phys.\  {\bf {42}} (1980) 97.
\bibitem{BL}
  G.P.\ Lepage and S.J.\ Brodsky, Phys.\ Lett.\ B {\bf 87} (1979) 359;
  Phys.\ Rev.\ D {\bf 22} (1980) 2157, Phys. Rev. {\bf D24} (1981) 1808.
\bibitem{soft}
N.~Isgur and C.H.~Llewellyn Smith,
Phys. Lett. {\bf B217} (1989) 535;\\
A.V.~Radyushkin,
Nucl. Phys. {\bf A527} (1991) 153C;
{\it ibid}. {\bf A532} (1991) 141.

\bibitem{LCSR}
 I.I.\ Balitsky, V.M.\ Braun and A.V.\ Kolesnichenko,
  Nucl.\ Phys.\ {\bf B312} (1989) 509;
 V.M.\ Braun and I.E.\ Filyanov, Z.\ Phys.\ {\bf {C 44}}
  (1989) 157;\\
V.L.\ Chernyak and I.R.\ Zhitnitskii, Nucl.\ Phys.\ {\bf B345} (1990) 137.


\bibitem{SVZ}
M.A.~Shifman, A.I.~Vainshtein and V.I.~Zakharov,
Nucl. Phys. {\bf B147} (1979) 385.


\bibitem{BH94}
V.M.~Braun and I.~Halperin,
Phys. Lett. {\bf B328} 457 (1994) 457 .

\bibitem{BKM}
V.M.~Braun, A.~Khodjamirian and M.~Maul,
hep-ph/9907495.

\bibitem{AK}
A.~Khodjamirian,
Eur. Phys. J. {\bf C6} (1999) 477.



\bibitem{IofSmilRad} 
B.L. Ioffe and A.V. Smilga, Phys. Lett. {\bf 114B} (1982) 353; 
Nucl. Phys. {\bf B216} (1983) 373;\\ 
V.A. Nesterenko and A.V. Radyushkin, Phys. Lett. {\bf 115B} (1982) 410. 




\bibitem{exp}
C.J.~Bebek {\it et al.},
Phys. Rev. {\bf D17} (1978) 1693;
S.R.~Amendolia {\it et al.}
[NA7 Collaboration],
Nucl. Phys. {\bf B277} (1986) 168.



\bibitem{GIKO}
A.S.~Gorsky, B.L.~Ioffe, A.Yu.~Khodjamirian and A.G.~Oganesian,
Z.\ Phys.\ {\bf C44} (1989) 523.



\bibitem{CELLO} CELLO Collaboration (H.-J. Behrend et. al),
Z. Phys. {\bf C49} (1991) 401; CLEO Collaboration (J. Gronberg et al.),
Phys. Rev. {\bf D57} (1998) 33. 


\bibitem{SY}
A.~Schmedding and O.~Yakovlev,
hep-ph/9905392.


\bibitem{heavy} 
V.M. Belyaev, A. Khodjamirian and  R. R\"uckl, Z. Phys. {\bf C 60} (1993) 349;  
V.M.~Belyaev, V.M.~Braun, A.~Khodjamirian and R.~R\"uckl,
Phys. Rev. {\bf D51} (1995) 6177;\\
P.~Ball and V.M.~Braun, Phys. Rev. {\bf D55} (1997) 5561;
Phys.\ Rev.\ {\bf D58} (1998) 094016;\\
A. Khodjamirian and R. R\"uckl, hep-ph/9801443, in 
Heavy Flavours II, eds. A.J. Buras and M. Lindner 
(World Scientific, Singapore,1998), p.345;\\
P.~Ball, JHEP {\bf 09} (1998) 005.

\bibitem{Bagan}
E.~Bagan, P.~Ball and V.M.~Braun,
Phys.\ Lett.\ {\bf B417} (1998) 154.



\end{thebibliography}
\end{document}